\documentclass{rnaastex}

\begin{document}

\title{Pole, pericenter and nodes of the interstellar minor body A/2017~U1}

%% Note that the corresponding author command and emails has to come
%% before everything else. Also place all the emails in the \email
%% command instead of using multiple \email calls.
\correspondingauthor{Carlos~de~la~Fuente~Marcos}
\email{nbplanet@ucm.es}

\author[0000-0003-3894-8609]{Carlos~de~la~Fuente~Marcos}
\affiliation{Universidad Complutense de Madrid \\
             Ciudad Universitaria, E-28040 Madrid, Spain}

\author[0000-0002-5319-5716]{Ra\'ul~de~la~Fuente~Marcos}
\affiliation{Universidad Complutense de Madrid \\
             Ciudad Universitaria, E-28040 Madrid, Spain}

%% Note that RNAAS manuscripts DO NOT have abstracts.
%% See the online documentation for the full list of available subject
%% keywords and the rules for their use.
\keywords{minor planets, asteroids: general --- minor planets, asteroids: individual (A/2017~U1) ---
comets: general}

%% Start the main body of the article. If no sections in the 
%% research note leave the \section call blank to make the title.
\section{} 

Interstellar minor bodies can be identified by their positive barycentric energy (i.e., they follow hyperbolic paths as their orbital eccentricities, 
$e$, are $>1$) and/or by their unusual chemical composition even if their $e<1$ ---e.g., the comets C/1996 B2 (Hyakutake) \citep{1996Sci...272.1310M} 
or 96P/Machholz~1 \citep{2007ApJ...664L.119L,2008AJ....136.2204S}. They can come from deep space or be produced locally, after close encounters with 
the planets or the Sun ---e.g., comet C/1980~E1 (Bowell) was ejected from the Solar System with $e=1.0577$ after a flyby with Jupiter 
\citep{1982M&P....26..311B,2009Icar..201..719M,2013RMxAA..49..111B}. Captured interstellar minor bodies can be returned to deep space after flybys 
---e.g., comet 96P/Machholz~1 may be ejected in the relatively near future \citep{2015MNRAS.446.1867D}. Interstellar minor bodies represent a rare and 
unique opportunity to have direct access to material from beyond the Solar System and the likelihood of observing them has been studied thoroughly 
\citep{1989ApJ...346L.105M,1992A&A...259..682K,1993A&A...275..298S,2009ApJ...704..733M,2011AJ....141..155J,2016ApJ...825...51C,2017AJ....153..133E}.

Here, we explore some peculiar orbital features of the recently discovered asteroid A/2017~U1 
\citep{A/2017U1}\footnote{\href{http://www.minorplanetcenter.net/mpec/K17/K17UI3.html}{M.P.E.C. 2017-U183}} that may help in understanding its 
nature. The discussion presented here is based on data publicly available from \href{http://ssd.jpl.nasa.gov/sbdb.cgi}{JPL's Small-Body Database}, 
but the results are novel and have not appeared elsewhere. Its heliocentric orbit determination (barycentric values in parentheses) as of 
2017-Oct-28 is: perihelion distance, $q$ = 0.254$\pm$0.002~au (0.257~au), $e$ = 1.196$\pm$0.004 (1.199), inclination, $i$ = 122\fdg6$\pm$0\fdg2 
(122\fdg9), longitude of the ascending node, $\Omega$ = 24\fdg605$\pm$0\fdg007 (24\fdg791), and argument of perihelion, $\omega$ = 241\fdg5$\pm$0\fdg3 
(241\fdg8);\footnote{Epoch 2458049.5 (2017-Oct-23.0) TDB} this solution is based on 59 observations for a data-arc span of 12 days. The number of 
known hyperbolic comets stands at 333 and there are five objects with $e>1.01$ ---C/1947~F1 (Rondanina-Bester), $e=1.0165$, C/1980~E1, $e=1.0577$, 
C/1997~P2 (Spacewatch), $e=1.0279$, C/1999~U2 (SOHO), $e=1.0162$, and C/2008~J4 (McNaught), $e=1.0279$. A/2017~U1 is a clear outlier, at the 
$\sim$49$\sigma$ level, when considering the average value of the eccentricity of known hyperbolic comets, 1.002$\pm$0.004. Some, perhaps all, the 
comets with $e>1.01$ have been ejected after close encounters with the Sun or the planets. 

\begin{figure}[!ht]
\begin{center}
\includegraphics[scale=0.6,angle=0]{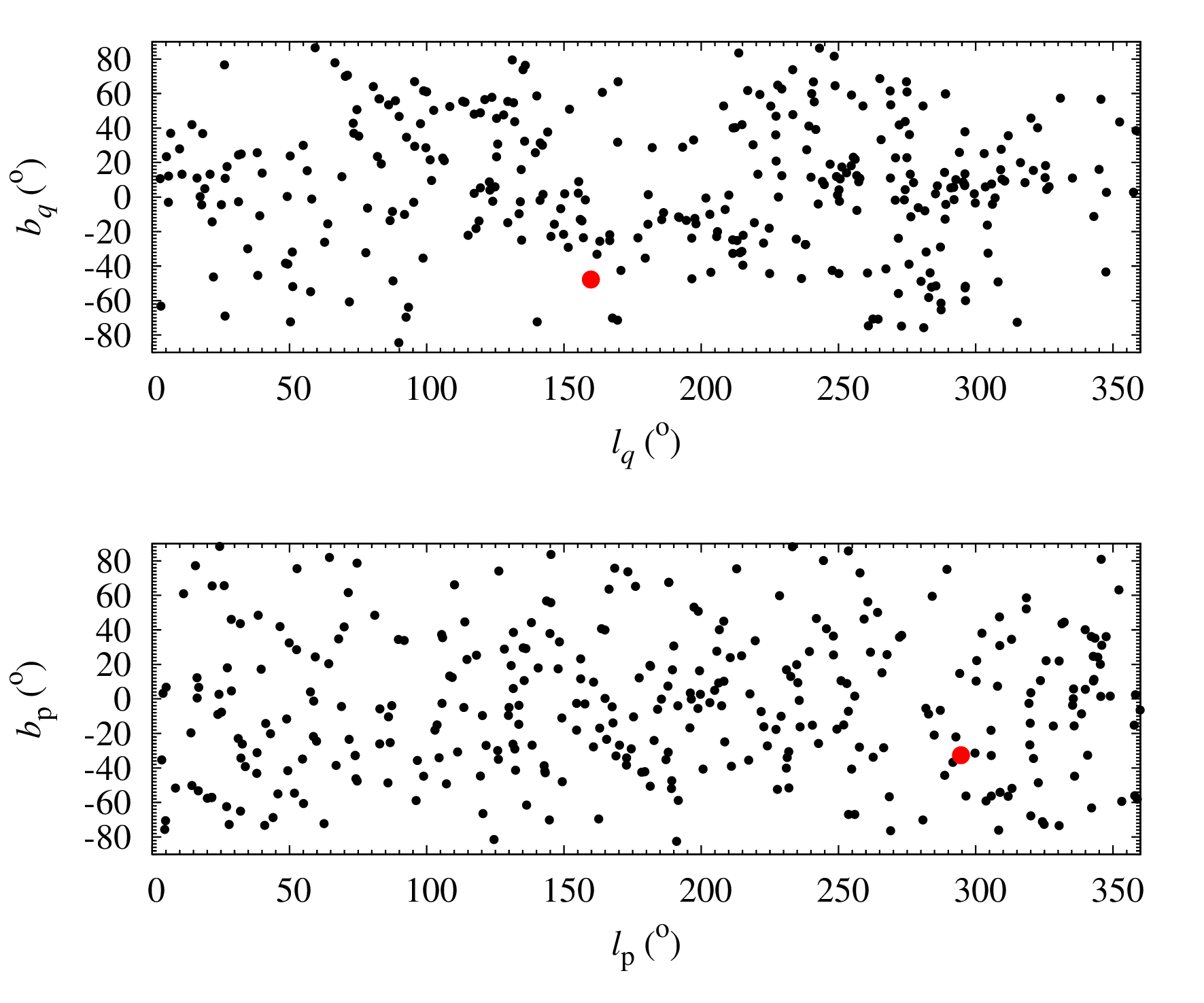}
\includegraphics[scale=0.4,angle=0]{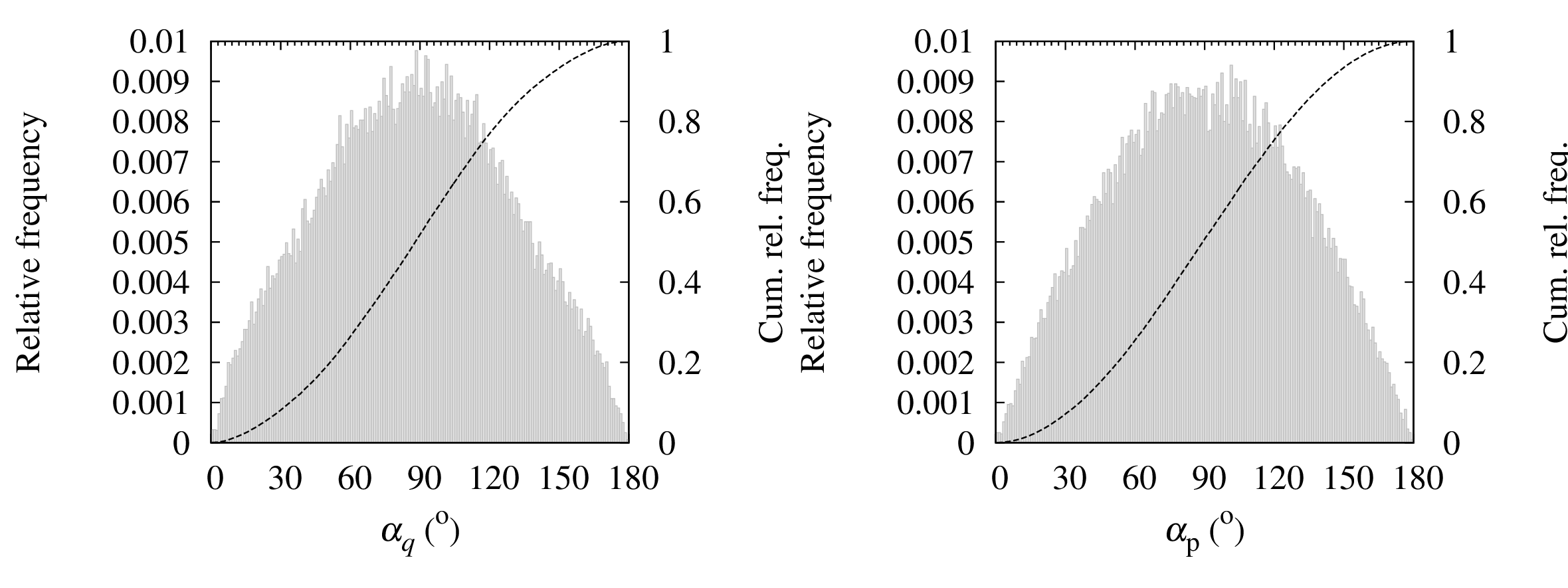}
\caption{Pericenters (top panel) and poles (middle panel) of the known hyperbolic minor bodies (334); A/2017 U1 is plotted in red.
         Distributions of possible angular separations between pericenters (bottom, left-hand side panel) and poles (bottom, right-hand side 
         panel) of the same sample.  
\label{fig:1}}
\end{center}
\end{figure}

Regarding the orientation of its orbit in space, the positions of pericenters ---ecliptic coordinates at pericenter, $(l_q, b_q)$--- and poles 
---projected pole positions, $(l_{\rm p}, b_{\rm p})$--- of known hyperbolic comets computed as described by \citet{2016MNRAS.462.1972D} are 
displayed in Figure~\ref{fig:1}, top and middle panels. The orientation of the orbital plane of A/2017 U1 seems to be away from any 
obvious clusters (real or induced by observational biases). The closest objects in terms of polar separation, $\alpha_{\rm p}$, are C/1970~N1 (Abe) 
(4\fdg79) and C/2014~A4 (SONEAR) (4\fdg50). When considering both $\alpha_{\rm p}$ and pericenter separations, $\alpha_q$, the smallest
relative angular separations are $>20\degr$. The distributions of $\alpha_q$ and $\alpha_{\rm p}$ of all the pairs in the sample are plotted
in the bottom panels. The hyperbolic comets with the closest mutual orbital orientation are the pieces of C/1986~P1 (Wilson), C/1996~J1-A 
and C/1996~J1-B (Evans-Drinkwater), and C/2011~J2 and C/2011~J2-B (LINEAR), all of them fragmented comets. Some of the clustering visible in
Figure~\ref{fig:1}, top and middle panels, may be due to dynamically correlated objects, probably the result of break-ups. The descending node of 
A/2017~U1 is located at 0.355~au and the ascending node at 1.301~au, well away from the paths of the planets of the Solar System and the Sun. 

All these orbital properties appear to confirm A/2017 U1 as the first known interstellar asteroid (either truly extrasolar or, far less likely, a 
returning former Solar System minor body that was ejected very long ago). 

%% An example figure call using \includegraphics

%% An example table using AASTeX's deluxetable. Note that since
%% only one figure OR one table is allowed this is commented out.
%\begin{deluxetable}{ccl}
%\tablecaption{Example table some English and Greek letters\label{tab:1}}
%\tablehead{
%\colhead{Index number} & \colhead{English} & \colhead{Greek}
%}
%\startdata
%1 & a & alpha ($\alpha$) \\
%2 & b & beta ($\beta$) \\
%3 & c & gamma ($\gamma$) \\
%4 & d & delta ($\delta$) \\
%5 & e & epsilon ($\epsilon$) \\
%\enddata
%\tablecomments{Long tables should only show a short example with the full
%version as a machine readable table with the article.}
%\end{deluxetable}  

\acknowledgments

We thank A.~I. G\'omez de Castro, I. Lizasoain and L. Hern\'andez Y\'a\~nez of the Universidad Complutense de Madrid (UCM) for providing access 
to computing facilities. This work was partially supported by the Spanish `Ministerio de Econom\'{\i}a y Competitividad' (MINECO) under grant 
ESP2014-54243-R. Part of the calculations and the data analysis were completed on the EOLO cluster of the UCM. EOLO, the HPC of Climate Change of 
the International Campus of Excellence of Moncloa, is funded by the MECD and MICINN. This is a contribution to the CEI Moncloa. In preparation of 
this Note, we made use of the NASA Astrophysics Data System and the MPC data server.


\begin{thebibliography}{}

\bibitem[Bacci et al.(2017)]{A/2017U1} Bacci P., Maestripieri M., Tesi L., et al.\ 2017, Minor Planet Electronic Circulars, 2017-U183 

\bibitem[Branham(2013)]{2013RMxAA..49..111B} Branham, R.~L., Jr.\ 2013, \rmxaa, 49, 111

\bibitem[Buffoni et al.(1982)]{1982M&P....26..311B} Buffoni, L., Scardia, M., \& Manara, A.\ 1982, Moon and Planets, 26, 311

\bibitem[Cook et al.(2016)]{2016ApJ...825...51C} Cook, N.~V., Ragozzine, D., Granvik, M., \& Stephens, D.~C.\ 2016, \apj, 825, 51

\bibitem[de la Fuente Marcos \& de la Fuente Marcos(2016)]{2016MNRAS.462.1972D} de la Fuente Marcos, C., \& de la Fuente Marcos, R.\ 2016, \mnras, 462, 1972

\bibitem[de la Fuente Marcos et al.(2015)]{2015MNRAS.446.1867D} de la Fuente Marcos, C., de la Fuente Marcos, R., \& Aarseth, S.~J.\ 2015, \mnras, 446, 1867

\bibitem[Engelhardt et al.(2017)]{2017AJ....153..133E} Engelhardt, T., Jedicke, R., Vere{\v s}, P., et al.\ 2017, \aj, 153, 133

\bibitem[Jura(2011)]{2011AJ....141..155J} Jura, M.\ 2011, \aj, 141, 155

\bibitem[Kresak(1992)]{1992A&A...259..682K} Kresak, L.\ 1992, \aap, 259, 682

\bibitem[Langland-Shula \& Smith(2007)]{2007ApJ...664L.119L} Langland-Shula, L.~E., \& Smith, G.~H.\ 2007, \apjl, 664, L119

\bibitem[McGlynn \& Chapman(1989)]{1989ApJ...346L.105M} McGlynn, T.~A., \& Chapman, R.~D.\ 1989, \apjl, 346, L105

\bibitem[Meech et al.(2009)]{2009Icar..201..719M} Meech, K.~J., Pittichov{\'a}, J., Bar-Nun, A., et al.\ 2009, \icarus, 201, 719

\bibitem[Moro-Mart{\'{\i}}n et al.(2009)]{2009ApJ...704..733M} Moro-Mart{\'{\i}}n, A., Turner, E.~L., \& Loeb, A.\ 2009, \apj, 704, 733

\bibitem[Mumma et al.(1996)]{1996Sci...272.1310M} Mumma, M.~J., Disanti, M.~A., dello Russo, N., et al.\ 1996, Science, 272, 1310

\bibitem[Schleicher(2008)]{2008AJ....136.2204S} Schleicher, D.~G.\ 2008, \aj, 136, 2204

\bibitem[Sen \& Rama(1993)]{1993A&A...275..298S} Sen, A.~K., \& Rama, N.~C.\ 1993, \aap, 275, 298

\end{thebibliography}
\end{document}